\documentclass[reprint, amsmath,amssymb, aps,
]{revtex4-2}

\usepackage[normalem]{ulem}
\usepackage{ragged2e}
\usepackage{multirow}
\usepackage{array}
\usepackage{comment}
\usepackage{graphicx}
\usepackage{amsmath}
\usepackage{amssymb}
\usepackage[none]{hyphenat}
\usepackage{xcolor}
\usepackage{hyperref}  
\usepackage{natbib}
\usepackage{tcolorbox}
\usepackage{enumitem}
\usepackage{tikz}
\usepackage{tikz-3dplot}
\usetikzlibrary{positioning}
\usepgflibrary{shapes.geometric}
\hypersetup{colorlinks=true,urlcolor=blue,citecolor=blue,linkcolor=blue}

\begin{document}
\sloppy
\preprint{APS/123-QED}

\title{Sensemaking and Scientific Modeling: Intertwined processes analyzed in the context of physics problem solving}

\author{Amogh Sirnoorkar}
\affiliation{Department of Physics, Kansas State University, Manhattan, KS 66506}

\author{Paul D. O. Bergeron}
\affiliation{Department of Physics and Astronomy, Michigan State University\\ Department of Chemistry, Michigan State University}

\author{James T. Laverty}
\email{laverty@phys.ksu.edu}
\affiliation{Department of Physics, Kansas State University, Manhattan, KS 66506}

\date{\today}
\begin{abstract}

Researchers in physics education have advocated both for including modeling in science classrooms as well as promoting student engagement with sensemaking.
These two processes facilitate the generation of new knowledge by connecting to one's existing ideas. Despite being two distinct processes, modeling is often described as sensemaking of the physical world. In the current work, we provide an explicit, framework-based analysis of the intertwining between modeling and sensemaking by analyzing think-aloud interviews of two students solving a physics problem. While one student completes the task, the other abandons their approach. The case studies reveal that particular aspects of modeling and sensemaking processes co-occur. For instance, the priming on the `given' information from the problem statement constituted the students' engagement with their mental models, and their attempts to resolve inconsistencies in understanding involved the use of external representations. We find that barriers experienced in modeling can inhibit students' sustained sensemaking. 
These results suggest ways for future research to support students' sensemaking in physics by promoting modeling practices.

\end{abstract}
\maketitle

\section{Introduction}
\label{sec:intro}
Students' meaningful engagement with curriculum has been a key area of focus in physics education research (PER).  One thread of this endeavor focuses on students' use of knowledge from their lived experiences and from formal instruction to reason about their surrounding world. This form of reasoning, referred to as sensemaking~\cite{defining}, assists in generating new knowledge by leveraging one's existing ideas~\cite{gupta2011beyond,hutchison2010attending}. Sensemaking is also associated with the ways disciplinary experts engage in knowledge construction~\cite{berland2009making,ford2012dialogic}. Given this significance, the field has experienced an increase in the investigations concerning the mechanisms~\cite{conceptualblending} and the nature of tasks~\cite{computation,kuo2020assessing,gifford2020curriculum,studentperspectives,methodssirnoorkar} associated with sensemaking. In addition to investigating the process, its significance, and ways of promoting it, researchers have also noted the close association sensemaking shares with modeling~\cite{defining,chen2022epistemic,russ2017intertwining,schwarz2009developing,schwarz2017helping,k-12framework,passmore2014models,sands2021modeling}.

Modeling is a common cognitive process through which humans comprehend their surrounding world~\cite{passmore2014models,sands2021modeling}. When modeling, the cognitive agent (the person engaging in modeling) abstracts and simplifies the target system to facilitate an explanation or a prediction~\cite{schwarz2009developing,ddi,dfdif,bergeron2021towards}. During this process, ideas from one's knowledge are organized and applied to reason about various phenomena. Consequently, model-based reasoning can be a crucial component of sensemaking~\cite{russ2017intertwining}. Several studies have noted that modeling and sensemaking share several common features including the common objective `to figure it out'~\cite{defining}. The two processes tend to consist of multiple phases including priming of prior knowledge of the concerned context, noticing discrepancies while reasoning, and generating new knowledge by connecting one's existing ideas~\cite{chen2022epistemic,sensemakinggame}.  

Furthermore, the connection between sensemaking and modeling has been described quite explicitly. For instance, Sands~\cite{sands2021modeling} proposes the `ACME protocol' elucidating mathematical modeling as the process of making sense of the physical world through three distinct stages. These include (i) Assessing the problem (making sense of the problem), (ii) Constructing the Model (making sure that the model makes sense), and (iii) Evaluating the model (making sense of the physical world using the model). These stages encompass the construction of a mental model as the modeler makes sense of the problem, expresses the ideas of the mental model mathematically (particularly during problem solving), and tests the model through its use to generate explanations or make predictions.

However, most of the above-mentioned accounts in the literature on the interplay between sensemaking and modeling focus on describing what modeling entails, but rarely articulate what constitutes making sense of something. An explicit description is important since sensemaking is a complex cognitive process in itself~\cite{defining,nokes2013toward}. A better understanding of how engaging with models facilitates sensemaking requires a detailed account of the sensemaking process. Additionally, we are not aware of any demonstration of the association between the components of the two processes through explicit frameworks. A nuanced description of the association between the corresponding elements can further the current understanding about the two widely studied processes. 

In the current work, we probe the association between the two processes by qualitatively analyzing the case studies of two students, Matthew and Ken, sensemaking about a physics problem by modeling the given context. We examine the students' modeling by noting their construction of mental models~\cite{grecaMentalPhysicalMathematical2002}, and their subsequent expression of the model's ideas using Su\'{a}rez's Denotative Function (DF), Demonstration (D), and Inferential Function (IF) -- the  DFDIF account of modeling~\cite{dfdif}. Additionally, we also examine the students' sensemaking through Odden and Russ's four stages of the Sensemaking Epistemic Game~\cite{sensemakinggame}. While Matthew completes the task (albeit with minor deviations from expected calculations), Ken opts to quit the problem-solving exercise abruptly. In Matthew's case, we note his construction of a mental model, and engagement with the DFDIF components to entail navigation through the stages of the Sensemaking Epistemic Game. Observations from Ken's case suggest that when the barriers to modeling are experienced, sensemaking is inhibited.

The manuscript is organized as follows: In the next section, we review the literature in PER on sensemaking and modeling. In Section~\ref{sec:theory}, we detail the DFDIF account of modeling and the Sensemaking Epistemic Game. We then discuss the data collection, context of the current study, and analysis of our data in Section~\ref{sec:methods}. In Sections~\ref{sec:Matthew-attempt} and~\ref{sec:ken-attempt}, we analyze the case studies through the described frameworks and discuss the observations in Section~\ref{sec:Discussion}. We end by discussing the affordances, limitations, and future explorations of this work in Section~\ref{sec:conclusion}.

\section{Literature Review}
\label{sec:lit review}

Sensemaking and modeling have been extensively studied in PER. An exhaustive account of the vast scholarly work on these two processes is beyond the scope of this work. However, a broad overview of the literature indicates that researchers have investigated these two constructs by defining them in a variety of ways. The context involved in these investigations range from students' reasoning of an everyday phenomenon to the use of mathematics in physics problem solving. In the following subsections, we will attempt to paint a picture of the broad trends in the literature as it relates to our current study. For a fuller account of the state of research into sensemaking and modeling, we recommend references~\cite{defining,zhao2021development} and references~\cite{handbook,gentner2014mental}, respectively. 

\subsection{Sensemaking}
Broadly, the literature on sensemaking can be classified into: (i) scientific and (ii) mathematical. These categories follow from studies focusing on students' qualitative reasoning on scientific phenomena and their quantitative reasoning during physics problem solving. Research into students' sensemaking on scientific phenomena  --  scientific sensemaking --  takes numerous approaches, including development of various theoretical perspectives of students' reasoning. Alternatively, research on mathematical sensemaking focuses  on investigating students' blending of formal mathematics with intuitive and conceptual arguments. In the following subsections, we describe these two sub-categories in detail. 

\subsubsection{Scientific sensemaking}

Literature on scientific sensemaking primarily consists of investigations probing students' sensemaking of scientific phenomenon. These studies have considered diverse descriptions of what accounts as sensemaking. Noting the lack of a unitary account, Odden and Russ~\cite{defining} identify three approaches through which sensemaking has been examined by the contemporary science education literature. These include: sensemaking as a stance towards science learning, as a cognitive process, and as a discourse practice.  

As a stance towards science learning, sensemaking has been considered from the perspective of epistemological framing -- a tacit understanding of `what's going on here'~\cite{hammer2005resources,sirnoorkar2020qualitative}. In a sensemaking epistemological frame, students generate novel explanations from lived experiences to address a noticed inconsistency in their understanding~\cite{hutchison2010attending,zohrabi2020processes,rosenberg2006multiple,danielak2014marginalized}. The second approach considers sensemaking as a cognitive process involving modeling of contexts through coordination of multiple representations~\cite{nokes2013toward,chiu2014supporting,linn1995designing,shen2011technology}. And the last approach -- as a discourse practice -- deals with construction and critique of generated explanations in response to a noticed gap in one's understanding~\cite{ford2012dialogic,berland2009making}. 

Odden and Russ~\cite{defining} weave these threads into a common strand, proposing an overarching definition of sensemaking as a dynamic process whose objective is `to figure something out'. This process is often motivated towards ascertaining the underlying mechanism of an observed phenomenon (Refer Section~\ref{subsec:sensemaking-theory} and Reference~\cite{defining} for more details). Numerous contemporary explorations have adopted this definition to advance the understanding of students' sensemaking~\cite{odden2021conceptual,sensemakinggame,chemsensemaking,odden2019vexing}. 

\subsubsection{Mathematical sensemaking}

The second category of the sensemaking literature -- mathematical sensemaking -- primarily involves investigations focusing on blending mathematics with conceptual physics during problem solving. Similar to its scientific counterpart, mathematical sensemaking has been defined in diverse ways, including ``\textit{leveraging coherence between formal mathematics and conceptual understanding}''~\cite{kuo2020assessing}, ``\textit{looking for coherence between the structure of the mathematical formalism and causal or functional relations in the world}''~\cite{dreyfus2017mathematical}, and ``\textit{a subset of sense making activities that privilege the use of mathematical formalisms in generating an explanation}''~\cite{gifford2020categorical}. Researchers have modeled mathematical sensemaking through various constructs such as `symbolic forms'~\cite{sherin2001students,dreyfus2017mathematical,zohrabi2020processes}, `blended processing'~\cite{kuo2013students}, `Calculation-Concept crossover' paradigm~\cite{kuo2020assessing}, mediated cognition~\cite{gifford2020categorical,inproceedings} and `blended sensemaking'~\cite{zhao2021development}. Through these constructs, the concerned studies look into how students express and extract meaning in physics through mathematical formalisms, coordinate among multiple representations, and employ quantitative reasoning to obtain qualitative insights and vice versa. 

The current work draws elements from both scientific and mathematical sensemaking literature. Our data involves students blending mathematical arguments with conceptual insights (mathematical sensemaking) in order to make sense of a real-world context (scientific sensemaking). We analyze students' reasoning by adopting Odden and Russ' account of sensemaking~\cite{defining}. Refer~\ref{subsec:sensemaking-theory} for additional details.

\subsection{Modeling}

Model-based reasoning has gained considerable attention in PER especially in the last three decades. The increasing emphasis on modeling is evident from `Developing and Using Models' being considered as one of the key scientific practices to be promoted in classrooms~\cite{NGSS,k-12framework}. The literature on modeling in PER can be mainly classified into:  analyzing students' mental models~\cite{frederiksen1999dynamic,grecaMentalPhysicalMathematical2002,bao2006model,fazio2013investigating,korhasan2015influence,bekaert2022identifying}, modeling during problem solving~\cite{hestenes1987toward, halloun1996schematic,pawl2009modeling,ACER,redish2008looking}, promoting modeling in classroom instruction~\cite{hestenes1992modeling,hestenes1987toward,etkina2006role,wells1995modeling,sawtelle2010positive,brewe2010toward,brewe2008modeling,halloun1987modeling,brewe2009modeling,brewe2013extending,halloun1996schematic,mcpadden2017impact}, and modeling in laboratories~\cite{dounas2018modelling,koponen2007models,zwickl2015model,dounas2016investigating,rios2019using,stanley2017using,zwickl2014incorporating,dounas2018modelling,koponen2007models}.

Investigations into mental modeling have primarily focused on students' construction and deployment of mental models. Greca and Moreira~\cite{grecaMentalPhysicalMathematical2002} define mental models to be {\em `an internal representation which acts as a structural analogue of situations or processes'}. Researchers have observed students to possess several simultaneous mental models whose probability of deployment depends on various factors including student's mental state, assessment statements, instructional methodologies, and peer interactions~\cite{bao2006model,korhasan2015influence}.

Problem solving has been one of the primary contexts in which researchers have discussed modeling in physics education~\cite{hestenes1987toward, halloun1996schematic,pawl2009modeling,ACER,redish2008looking}. These studies have considered modeling in general, and mathematical modeling in particular to be an integral part of problem solving in physics. Researchers have noted several factors such as description of the target system, and the interactions between the system's components to guide the choice and construction of appropriate models during problem solving~\cite{pawl2009modeling,redish2008looking}.  

The third domain of investigations on modeling -- promoting modeling in classrooms (part of which overlaps with the studies on modeling during problem solving) -- has been advocated to emphasize the application of physics theories and principles to real-world contexts~\cite{hestenes1987toward,halloun1996schematic}. To promote this practice through teaching, models have been classified as mathematical models~\cite{hestenes1987toward,redish2008looking}, physical models~\cite{grecaMentalPhysicalMathematical2002}, conceptual models~\cite{halloun1996schematic}, and models of objects, systems, interactions and processes~\cite{etkina2006role}. Brewe and colleagues have extended the Modeling Instruction to university level by explicitly focusing on representations during classroom discourses~\cite{sawtelle2010positive,mcpadden2017impact,brewe2008modeling,brewe2009modeling,brewe2010toward,brewe2013extending}. This explicit emphasis on modeling has been found to positively impact self-efficacy of women in learning physics~\cite{sawtelle2010positive}, facilitate equitable learning environments for students~\cite{brewe2010toward}, cause positive attitudinal shifts towards physics learning~\cite{brewe2009modeling,brewe2013extending} and promote students' use of representations during problem solving~\cite{mcpadden2017impact}.


In addition to promoting modeling in classrooms, investigations in PER have also emphasized modeling during laboratory experiments. Reasoning around models in laboratories has been advocated since models act as the connecting link between theory and experimentation~\cite{koponen2007models}. The call for designing model-centered laboratory activities has been addressed through development of frameworks and assessments~\cite{dounas2018modelling,zwickl2015model,rios2019using}. Zwickl {\em et al.}~\cite{zwickl2014incorporating} have noted that an explicit emphasis on models in laboratories can enhance the  blending of conceptual and quantitative reasoning during experimentation.

In the landscape of modeling literature in PER, we aim to place this study in the categories of students' construction of mental models and modeling during problem solving. In the sections to follow, we discuss two case studies of students solving a physics problem by initially constructing a mental model and subsequently reasoning by constructing representations in order to make sense of the given task.  

\section{Theoretical frameworks}
\label{sec:theory}
As noted in Section~\ref{sec:intro}, the contemporary descriptions on the intertwining between modeling and sensemaking lack  (i) a detailed account of sensemaking, and (ii) a framework-based methodological approach in unpacking the intertwining between the two processes. We address the first concern by discussing the sensemaking process through the lens of the Sensemaking Epistemic Game~\cite{sensemakinggame} which succinctly describes how the sensemaking process begins, proceeds, and terminates through four distinct stages. On the other hand, we adopt the DFDIF account~\cite{dfdif}, which describes the modeling process in terms of three components thereby facilitating a categorical analysis of the association between modeling and sensemaking (thus addressing the second concern).

Below, we describe our adopted frameworks on sensemaking and modeling in addition to noting their relation with Sands' `ACME protocol'~\cite{sands2021modeling}, which is the most explicit discussion about the intertwining between the two processes. As noted in Section~\ref{sec:intro}, the ACME protocol describes mathematical modeling through three stages: (i) Making sense of the problem - developing a qualitative mental model about the target system, (ii) Making sure that the model makes sense - translating the qualitative ideas into mathematical relationships, and (iii) Making sense of the physical world using the model - physically interpreting the established mathematical relationships.

\subsection{Sensemaking Epistemic Game}
\label{subsec:sensemaking-theory}

\renewcommand{\arraystretch}{1.3}
\begin{table*}
\begin{center}
\caption{A brief description of the stages in the Sensemaking Epistemic Game. Stages 1-3 correspond to sensemaking.\label{tab:sensemaking}}

\begin{ruledtabular}

\begin{tabular}{ p{0.6cm} p{6.3cm} p{10.7cm}}

& Stages of Sensemaking Epistemic Game  & Description \\
\hline

& 0. Assembling of a knowledge framework & Prior to sensemaking, students prime their existing knowledge on the concerned task. \\

 & & \\

\multirow{4}{*}{\rotatebox[origin=c]{90}{Sensemaking \hspace{0.3cm}}}  & 1. Noticing a gap in knowledge &  Students transition into sensemaking by noticing an inconsistency between existing knowledge and the knowledge required from the task. \\

& 2. Generating an explanation   & Explanations are generated in response to the noticed inconsistency by connecting one's existing ideas. \\

& 3. Resolution  & Students conclude the sensemaking process upon generating a satisfactory explanation. \\

\end{tabular}
\end{ruledtabular}
\end{center}
\end{table*}

In the current work, we adopt Odden and Russ's following account of sensemaking as:
\begin{quote}
{\em a dynamic process of building or revising an explanation in order to `figure something out' - to ascertain the mechanism underlying a phenomenon in order to resolve a gap or inconsistency in one's understanding. One builds this explanation out of a mix of everyday knowledge and formal knowledge by iteratively proposing and connecting up different ideas on the subject. One also simultaneously checks that those connections and ideas are coherent, both with one another and with other ideas in one's knowledge system~\cite{defining}.}
\end{quote}

Odden and Russ also propose the Sensemaking Epistemic Game~\cite{sensemakinggame} framework to describe how the process begins, progresses, and concludes. This framework adopts the construct of epistemic games~\cite{Ferguson} to characterize the trajectory of sensemaking.  Epistemic games correspond to the set of rules employed when undertaking a scientific inquiry towards the goal of the task called a {\em target epistemic form}. These epistemic forms vary with the task under investigation. When playing any game, we abide by a certain set of rules which are referred to as the {\em constraints} of the game. 

Along with the target epistemic forms and constraints, epistemic games also have {\em entry conditions, moves, and exit conditions}. The circumstances around which a person begins to play the game are called {\em entry conditions}.  These conditions are often triggered by the nature of the inquiry. {\em Moves} are the set of actions taken at various stages of the game, and {\em exit conditions} represent the circumstances in which the inquiry is terminated. 

As an example, consider determining the displacement between the given points $A$ and $B$ in a two-dimensional Euclidean space. Let the points be defined by the coordinates $A \ (x=2, y=3)$ and $B \ (x=5, y=1)$. The {\em target epistemic form} of this inquiry (epistemic game) would correspond to ascertaining the magnitude and direction of the displacement vector defined between the points. The {\em constraint} in such a case would be to obtain a single numerical value along with its direction. The {\em entry and exit conditions}, that is, how one initiates and terminates the approach, can vary according to whether the prompt is open-ended or a multiple choice. Performing algebraic calculations on the given data or manually plotting the points on a graph would count as the {\em moves} of this game.

Odden and Russ adopt the epistemic games construct and describe the trajectory of sensemaking through the following four stages of the Sensemaking Epistemic Game~\cite{sensemakinggame}:
 
\begin{enumerate}[align=left]
\item[{\bf Stage 0:}] {\bf Assembling a knowledge framework}
\end{enumerate}   
Before sensemaking, students prime their existing knowledge on the given domain. This can include recalling the general understanding of the task-related concepts, priming on the provided information, etc.\ This stage of assembling prior knowledge is highly dependent on the activity's context and forms the precursor to sensemaking. 

\begin{enumerate}[align=left]
\item[{\bf Stage 1:}] {\bf Noticing a gap or inconsistency}
\end{enumerate}
Students notice a gap between their existing knowledge and the knowledge expected from the task. The noticing of the gap in one's knowledge system marks the entry condition into the epistemic game. This is often accompanied with recurring pauses, articulation of vexing questions, etc.\ 

\begin{enumerate}[align=left]
\item[{\bf Stage 2}] {\bf Generating an explanation:}
\end{enumerate}
In response to the noticed inconsistency, explanations are generated by connecting one's existing ideas and seeking coherence between them. These explanations are often drawn from the knowledge gained through one's lived  experiences and formal instruction.

\begin{enumerate}[align=left]
\item[{\bf Stage 3}] {\bf Resolution:}
\end{enumerate}    
Students reach the target epistemic form (goal of the task) upon generating a satisfactory explanation which addresses the noticed inconsistency. Resolution of the inconsistency marks the exit condition from the Sensemaking Epistemic Game. This stage is highlighted by confident articulation of the explanation or a claim with appropriate justification.  

Table~\ref{tab:sensemaking} summarizes the four stages of the Sensemaking Epistemic Game. While this framework is consistent with our adopted definition of sensemaking, it however differs from the Sands' `ACME protocol'~\cite{sands2021modeling}. The difference mainly lies in the meta-cognitive feature of noticing discrepancies while reasoning. Unlike the ACME protocol, the objective of sensemaking (according the Sensemaking Epistemic Game) is to address a perceived discrepancy in one's knowledge system.

In the sections to follow, we analyze students' sensemaking on a physics problem by categorizing their approach in terms of the above-mentioned stages of the Sensemaking Epistemic Game. In order to observe the intertwining between the components of sensemaking with modeling process, the same problem solving approaches are analyzed through the components of modeling described below.

\subsection{Modeling}

Similar to sensemaking, the notion of engaging with models has been a fragmented construct in terms of its diverse characterization. Models have typically been regarded as representations standing completely or partially in isomorphic relation to their target systems~\cite{giere1999using,chakravartty2010informational}. However, recent discussions in the philosophy of science literature have increasingly emphasized the role of human agent's cognitive interests in modeling a system~\cite{gouvea2017models,suarez2004inferential,handbook}. Consequently, in this work, we define modeling as

\begin{quote}
{\em an activity engaged in by a cognitive agent involving abstraction and simplification of a phenomenon in order to generate explanations or predictions}~\cite{schwarz2009developing,gouvea2017models}.    
\end{quote} 

In addition, consistent with the recent arguments in the literature, we consider modeling to involve an initial construction of a qualitative mental model, and subsequently expressing the model's ideas through external representations~\cite{sands2021modeling}. 

Mental models are {\em internal representations which act as structural analogue of situations or processes}~\cite{grecaMentalPhysicalMathematical2002} and have been noted as a precursor of constructing external representations~\cite{sands2021modeling}. Human beings' interaction with the physical world is mediated by mental models which are based on one's lived experiences and social interactions~\cite{johnson1983mental}. As representations of the physical world, the mental models are incomplete as they do not possess all the requisite information to comprehend and explain a complex context. Thus in the initial stages of a complex activity such as physics problem solving, mental modeling often involves making sense of the contextual information in light of one's existing knowledge~\cite{skovgaard2019cancellation,sands2021modeling}.

The external representations, on the other hand, often reflect the qualitative ideas contained in the mental model, and depict the specific features of the target system to be modeled. Suarez~\cite{dfdif}, building on the work of Hughes~\cite{ddi}, elucidates the process of modeling through external representations in terms of the following three components:

\begin{enumerate}[align=left] 
\item {\bf Denotative Function (DF):} 
\end{enumerate}    
Elements of the target system are specified by elements of the external representation when abstractly portraying a phenomenon. This feature of making a transition from the `physical world' to the `model world' marks this component of modeling.  

\begin{enumerate}[align=left]\setcounter{enumi}{1} 
\item {\bf Demonstration (D):}  
\end{enumerate}
The elucidation of the internal dynamics between the designated elements in the representation marks the second component of modeling. In physics, this is typically achieved by establishing mathematical relationships between the denoted elements of the representation. 

\begin{enumerate}[align=left]\setcounter{enumi}{2}
\item {\bf Inferential Function (IF):}
\end{enumerate} 
The theoretical relationships obtained in the Demonstration stage are then interpreted in terms of the target system. The reverse-transition from the `model world' to the physical world corresponds to the last component.

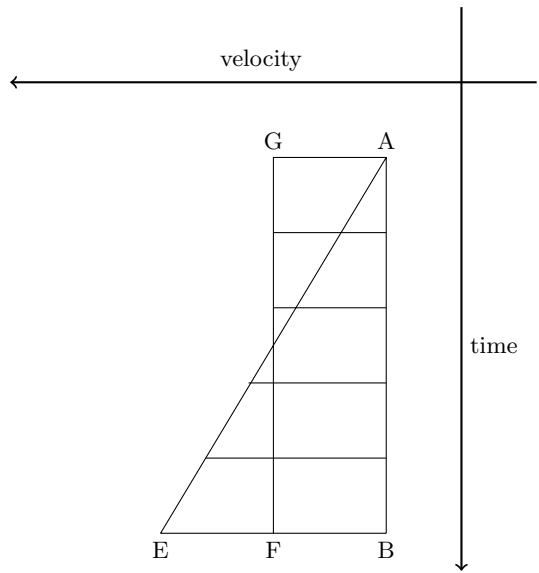
\begin{figure}
    \centering
\begin{tikzpicture}
\draw[->,thick] (6,6.5) -- (-1,6.5);
\draw[->,thick] (5,7.5) -- (5,0);
\draw (2.5,5.5) -- (2.5,0.5) (2.5,5.5) -- (4,5.5)  (2.5,0.5) -- (4,0.5) (4,5.5) -- (1,0.5) (1,0.5) -- (2.5,0.5) (4,5.5) -- (4,0.5);
\draw (2.5,1.5) -- (4,1.5) (2.5,2.5) -- (4,2.5) (2.5,3.5) -- (4,3.5) (2.5,4.5) -- (4,4.5) (2.5,1.5) -- (1.6,1.5) (2.5,2.5) -- (2.17,2.5);
\draw (5,3) node[right]{time} (3,6.8) node[left]{velocity};
\draw (2.5,5.5) node[above]{G} (4,5.5) node[above]{A} (1,0.5) node[below]{E} (2.5,0.5) node[below]{F} (4,0.5) node[below]{B};

\end{tikzpicture}
    \caption{Modified version of the Galileo's diagram relating the distances covered by a uniformly accelerating object starting from rest, to an object moving with constant speed. The length of vertical lines represent the time taken by the two objects to traverse a constant distance. The horizontal lines correspond to the increasing speeds. Refer~\cite{galilei1974two,ddi,dfdif} for the original diagram.}
\label{fig:galileo}
\end{figure}

Hughes~\cite{ddi} and Su\'{a}rez~\cite{dfdif} illustrate the above-mentioned components by referring to Galileo's kinematical problem introduced in the Third Day of his Discourses Concerning Two New Sciences~\cite{galilei1974two}. In Proposition I, Theorem I, Galileo demonstrates that two objects, one accelerating from rest and the other traveling at constant speed, would cover the same distance in a given time if the former's terminal speed is twice the latter's uniform speed. He elucidates this result by representing the kinematical scenario through a geometrical representation as shown in Figure~\ref{fig:galileo}. 

In the figure, the length of the line segment AB represents time taken by the accelerating object to cover a given distance (say $x$ units). AE is drawn such that increasing length of the horizontal lines from AB to AE represent the increasing degree of instantaneous speeds of the accelerating object. The parallelogram AGFB is constructed such that FG is parallel to AB and bisects EB. FG then represents the time taken by a uniformly moving object (as evidenced by the equidistant lines from AB) in traversing the distance $x$. Areas of the parallelogram AGFB and triangle AEB (i.e., the distances travelled by the two objects) can be shown to be equal provided the terminal speed of the accelerating object EB is twice the uniform speed FB of the other. 


\begin{table*}
\caption{A brief description of the DFDIF components of modeling.}
\label{tab:modeling}
\begin{ruledtabular}
\begin{tabular}{p{0.3\textwidth} p{0.6\textwidth}}

DFDIF Components of Modeling & Description \\
\colrule
Denotative Function (DF) & Elements of the target system are specified by elements of the representation when abstractly portraying a phenomenon. \\

Demonstration (D) & The elucidation of the internal dynamics between the denoted elements. \\

Inferential Function (IF) & The mapping of theoretical conclusions onto the target system.     \\
\end{tabular}
\end{ruledtabular}
\end{table*}  

The above proof is a clear example of modeling in which a physics problem is expressed as a problem in geometry. In order to abstractly express the kinematic scenario as a geometrical representation, the physical quantities (speed and time) are denoted as horizontal and vertical line segments. This feature of establishing a relationship between physical quantities and elements of the representation constitutes the Denotative Function component of modeling. Thereafter, areas of the parallelogram AGFB and triangle AEB are equated, with EB being twice EF. This elucidation of the internal dynamics between the denoted line segments marks the second component, i.e., Demonstration. Lastly, the geometrical results obtained in the Demonstration stage are interpreted in terms of kinematics reflecting the Inferential Function component.   

Our adopted DFDIF framework on modeling shares several common features with the Sands' `ACME' protocol~\cite{sands2021modeling}. The first stage of the protocol (assessing the problem) broadly aligns with the Denotative Function component wherein elements of the target system are identified. The second and the third stages (constructing and evaluating the model) reflect the remaining two components of the DFDIF framework wherein the identified elements are related (Demonstration) in order to generate explanations/predictions (Inferential Function).

\section{Research Questions}

In the rest of this paper, we seek to address following research questions: 

\begin{enumerate}
    \item How does engagement with mental models and the DFDIF components associate with navigation through the stages of the Sensemaking Epistemic Game?
    
    \item How does construction of mental models and engagement with DFDIF components influence students' sustained engagement in sensemaking?
\end{enumerate}

Answering the above research questions will further the current understanding of the intertwining between the elements of the two processes. A nuanced understanding of this intertwining also can assist in sustaining and promoting sensemaking through modeling in our learning environments.

\section{Methodology}
\label{sec:methods}
\subsection{Data Collection}

We adopt case study analysis to unpack the intertwining of modeling components with the stages of the Sensemaking Epistemic Game. This approach allowed us to deeply analyze the intertwining between the two processes through an in-depth analysis of a target phenomenon~\cite{creswell2016qualitative}. The case studies are derived from a pool of video data consisting of students from a Midwestern US university participating in think-aloud interviews~\cite{rios2019using}. The interviews were comprised of individual students (N=10) articulating their thoughts aloud while solving a set of physics problems. The interview protocol involved asking participants to treat the problem-solving exercise as an exam and were informed that the exercise was untimed and their participation had no bearing on their academic grades. Students were  compensated with \$20 for their participation. The interviewer interacted with participants only when asked, or to prompt them to express their thoughts aloud with questions such as ``{\em What are you thinking?''}. The participants were allowed to use calculators and were provided with an equation sheet along with the problem set. 

The problem set consisted of nine introductory college physics tasks designed to investigate students' engagement in Scientific Practices~\cite{k-12framework}. Of these nine problems, three (problems 7, 8 and 9) were specifically designed using the Three-Dimensional Learning Assessment Protocol (3D-LAP)~\cite{3DLAP} to elicit the Scientific Practice of `Developing and Using Models'~\cite{k-12framework}. In the rest of this paper, we focus on students' responses to problem 7 - the `Gravitron task' (Section~\ref{subsec:gravitron}). We focus on this particular task for two reasons. First, unlike the responses to the first six problems, a majority of students in our data reasoned about the Gravitron task by making arguments from their lived experiences and through physical gestures. This characteristic shift in approach in students' approaches caught our attention. Second, the Gravitron task is also the first 3D-LAP question in the problem set which students encountered and presumably engaged in the modeling practice.

Of the 10 participants, we focus on Matthew's and Ken's (both pseudonyms) approaches to the Gravitron problem. In addition to the audio/video clarity of their interviews, our choice to focus on the two students' approaches is informed by their clear articulation of arguments while solving the problem. The clarity of the arguments played a key role in identifying various elements of modeling and sensemaking processes. While both students sought to make sense of the Gravitron's scenario, they differed in their sustained engagement. Matthew's attempt involved successful navigation through the four stages of the epistemic game whereas Ken chose to quit the process abruptly. Comparison between these two case studies offer valuable insights on the circumstances in which students may opt to abandon sensemaking.

\subsection{The Gravitron Problem}
\label{subsec:gravitron}

The Gravitron task (see Figure~\ref{fig:Gravitron} for the problem statement) involves a rotating cylindrical amusement park ride in which the rider leans against the wall. With the given parameters, students are asked to determine whether riders would slide off the Gravitron's wall. 

\begin{figure}[tb]
    \centering
    \begin{tcolorbox}
     \justify{You are asked to design a Gravitron for the county fair, an amusement park ride where the  rider enters a hollow cylinder, radius of $4.6$~m, the rider leans against the wall and the room spins until it reaches angular velocity, at which point the floor lowers. The coefficient of static friction is $0.2$. You need this ride to sustain mass between $25$-$160$~kg to be able to ride safely and not slide off the wall. If the minimum $\omega$ is $3$~rad/s, will anyone slide down and off the wall at these masses? Explain your reasoning using diagrams, equations and words.}   
     \end{tcolorbox}
    \caption{Statement of the Gravitron problem}
    \label{fig:Gravitron}
\end{figure}

\begin{figure}
    \centering
\begin{tikzpicture}
\draw[dashed] (0,0) ellipse (1.6cm and 0.3cm);
\draw[dashed] (0,-4) ellipse (1.6cm and 0.3cm);
\draw[dashed] (-1.6,0)--(-1.6,-4);
\draw[dashed] (1.6,0)--(1.6,-4);
\filldraw[black] (1.6,-2) circle (3pt);
\draw [ultra thick,->] (1.6,-2)--(1.6,-1) ;
\draw [ultra thick,->] (1.6,-2)--(1.6,-3);
\draw [ultra thick,->] (1.6,-2)--(0.6,-2);
\node [right] at (1.7,-1) {Friction ($F_f$)};
\node [right] at (1.7,-3) {Gravity ($F_g$)};
\node [above] at (0,-1.8) {Normal force ($N$)};
\draw[thick, ->] (0.5,0.1) .. controls (-1.5,0.1) and (0,-0.2) .. (0.5,-0.1);
\end{tikzpicture}
    \caption{A force diagram representing the gravity, friction, and normal force acting on a rider in the drum of a Gravitron amusement park ride (see task statement in Figure~\ref{fig:Gravitron}). The dot in the diagram represents the rider's center of mass.}
    \label{fig:fbd}
\end{figure}
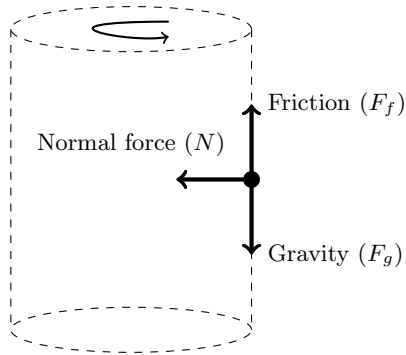

One of the ways to approach this task is by noting the forces acting on the rider through a force diagram as shown in Figure~\ref{fig:fbd}. If a rider is to be suspended without slipping, the vertically upward acting frictional force ($F_f$) offered by the wall must be at least equal to the downward pull of gravity ($F_g$). Mathematically, this argument can be expressed as:

\begin{equation}
    F_f = F_g \label{eq:1}
\end{equation}

Since the maximum friction offered by a surface is equal to the surface's coefficient of static friction ($\mu$) times the normal force ($N$), and as the normal force provides the necessary centripetal force, the above equation can be simplified as

\begin{equation}
\mu  (mr\omega^2) \geq mg   \label{eq:2}  
\end{equation}

where $m$ represents the rider's mass, $r \ (=4.6m) $ represents the Gravitron's radius, $\omega \ (=3 rad/s)$ represents the rider's angular velocity and $g \ (=9.8 m/s^2)$ represents the acceleration due to gravity. The simplification of Equation~\ref{eq:2} leads to the condition

\begin{equation}
    \mu r \omega^2 \geq g
\end{equation}

By substituting the given parameters, one observes that the above inequality does not hold true and consequently a rider would slide off the Gravitron's walls.

From the epistemic games perspective, the Gravitron task's {\em target epistemic form} corresponds to the final claim about the rider's status inside the Gravitron. This claim is further {\em constrained} by the two possibilities about the riders either falling off, or holding up against the Gravitron's walls.




\subsection{Data analysis}

After narrowing down the case studies, the first author iteratively viewed and transcribed Matthew's and Ken's interviews, taking into account the participants' speech, physical gestures, and written solutions. The transcript -- documented description of participants' attempts in terms of verbal arguments, gestures and written solutions --  was then examined and segmented into the stages of the Sensemaking Epistemic Game based on the discourse markers discussed in the literature~\cite{odden2019vexing,mathayas2019representational}. As evidence for the Noticing of an Inconsistency stage, we looked for instances of students `getting stuck', which were cued by markers such as articulation of puzzling questions in their arguments~\cite{odden2019vexing} accompanied by representational gesturing~\cite{mathayas2019representational}. Instances preceding this stage where students recalled or gathered task-related information marked the Assembling of a Knowledge Framework stage of sensemaking. The Generation of an Explanation stage was identified by noting the mathematical or conceptual arguments that the participants made after noticing inconsistencies in their understanding. Lastly, students' correct and/or complete articulation of their final claim marked the `Resolution' stage of the epistemic game. 

\begin{figure*}
    \centering
\includegraphics[scale=0.14]{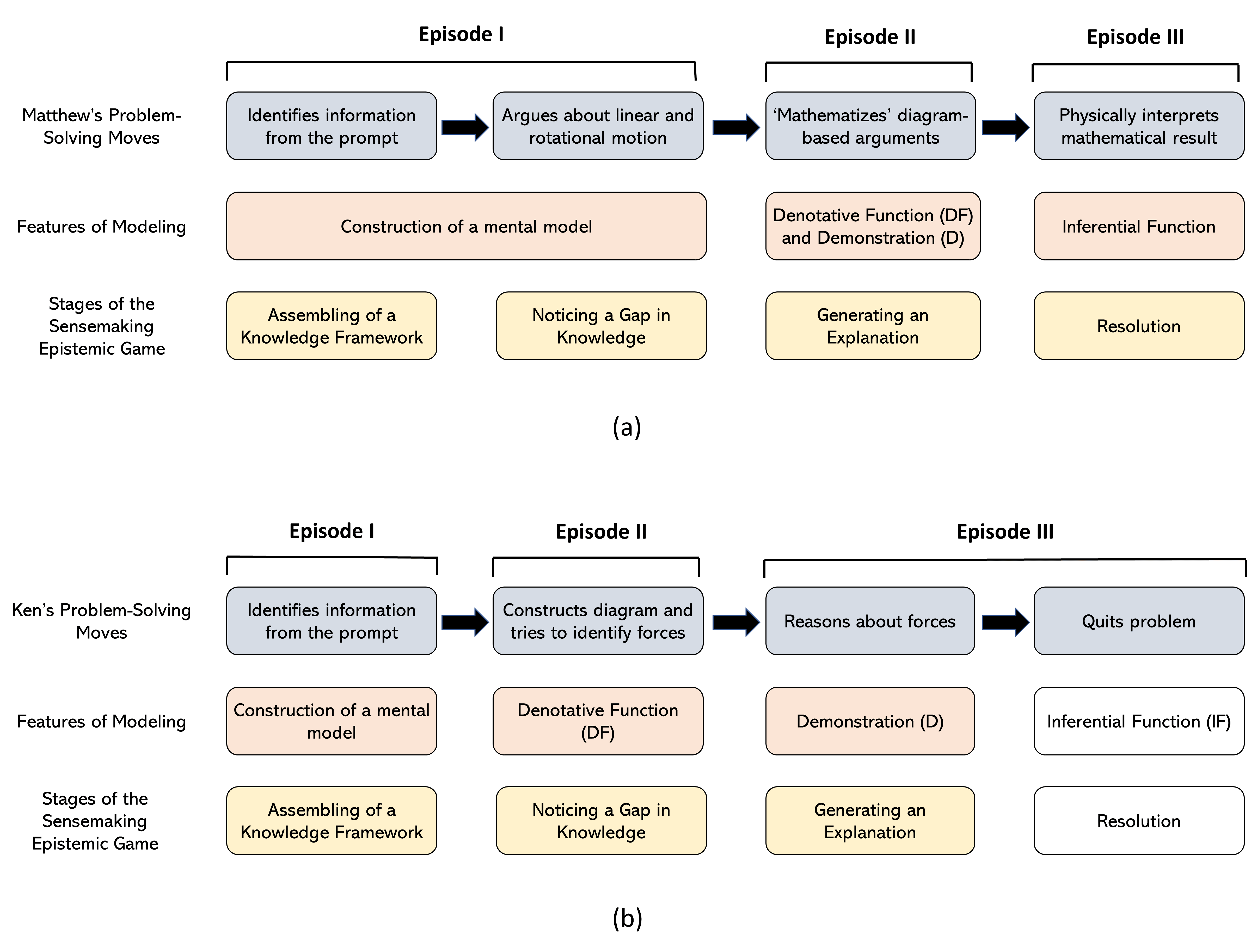}

    \caption{Alignment of Matthew's (a) and Ken's (b) problem-solving moves with the stages of the Sensemaking Epistemic Game and the features of modeling across the three episodes. In both the figures, the top rows describe the participants' problem-solving moves in their attempt at the Gravitron task, the middle row highlights their construction of a mental model and their engagement with the DFDIF components of modeling, and the bottom row identifies the stages of the Sensemaking Epistemic Game evidenced during their attempts.}
    \label{fig:participants-summary}
\end{figure*}

To capture the modeling component in the participants' work, the first author segmented the same transcript into two parts: before and after construction of an external representation. Students' arguments before construction of external representations, particularly those which accompanied representational gesturing~\cite{grecaMentalPhysicalMathematical2002,mathayas2019representational}, and those reflecting a sense of `incompleteness or uncertainty'~\cite{gentner2014mental} cued about their engagement with mental models. Note that the objective of the current work is neither to characterize nor analyze the participants' mental models, but rather to identify the instances of their mental modeling to make comparisons to the Sensemaking Epistemic Game.

We identified the modeling's DFDIF components by noting the participants' arguments during and after construction of external representations (diagrams or equations). For instance, associating forces with arrows in a force diagram, or designating an equation's algebraic symbols with Gravitron's parameters indicated the Denotative Function in the students' work. Highlighting the directional relationships between the denoted arrows, or establishing mathematical relationships between algebraic symbols reflected the Demonstration component. The physical interpretation of the relationships established in the Demonstration stage marked the modeling's Inferential Function. 

The segmented transcripts of the students' approaches in terms of stages of the epistemic game and components of modeling were then compared. The relative positioning of the modeling activities (mental modeling and engaging with DFDIF components) with the stages of the Sensemaking Epistemic Game highlighted the association between the two processes (Refer Figure~\ref{fig:participants-summary}).  In the following two sections, we provide a detailed account of Matthew's and Ken's approaches, categorized into modeling activities (construction of mental model and engaging with DFDIF components) and the stages of the Sensemaking Epistemic Game.

\section{Matthew's attempt at the Gravitron task}
\label{sec:Matthew-attempt}

\begin{figure*}
    \centering
    \includegraphics[scale=0.4]{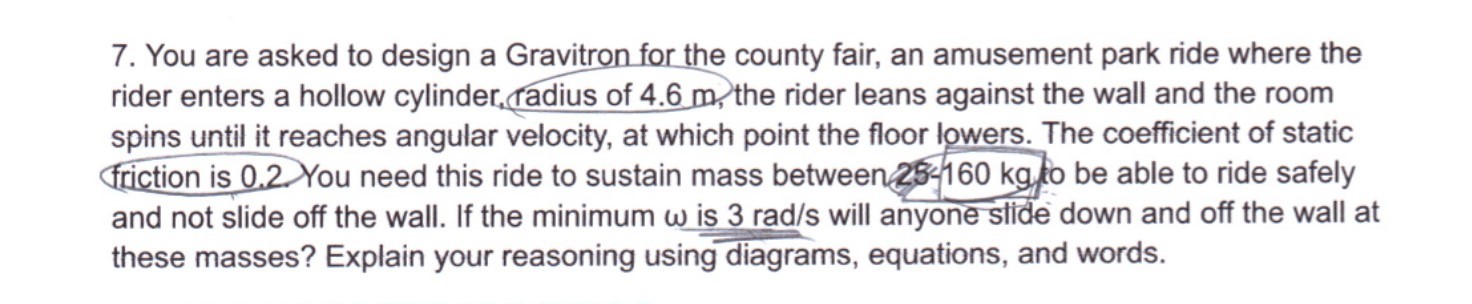}
    \includegraphics[scale=0.7]{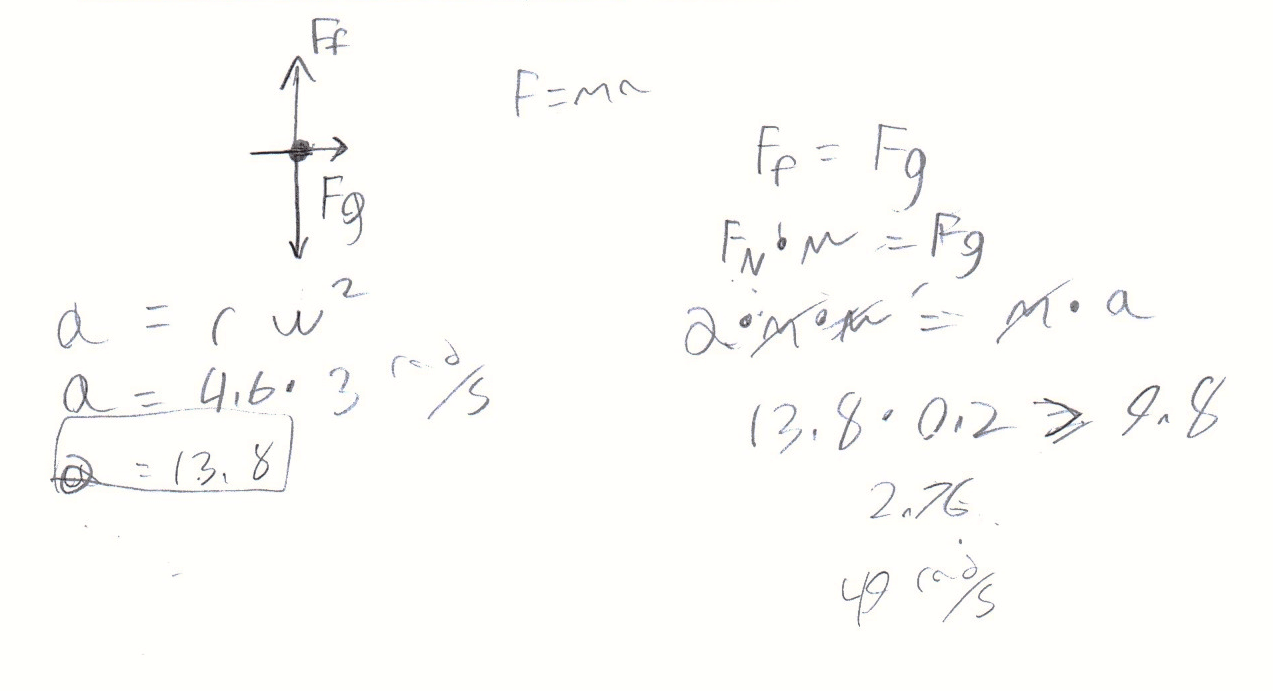}
    \caption{Matthew's written solution to the Gravitron problem. }
    \label{fig:Matthew-attempt}
\end{figure*}

Matthew begins by noting the given information before noticing a gap in his understanding. He addresses the perceived gap by constructing a force diagram, and translating the diagram-based arguments into an algebraic inequality. The validity of the inequality makes Matthew to conclude that riders would slide down the Gravitron's walls under the given conditions. 

\subsection{Episode I}
\label{subsec:matthew-episode1}
Matthew initiates by highlighting the parameters from the problem statement (as observed in Figure~\ref{fig:Matthew-attempt}) and says: 

\begin{quote}
    {\em So, I just went through and circled all the important information. Here it talks about the mass of the rider.}
\end{quote}

He then refers to the ease of free fall by a heavier person before shifting his argument to angular momentum. This argumentative shift is accompanied by a hand gesture and intermittent pauses. 

\begin{quote}
    {\em  If the person who weighs more, then its gonna be easier for them to drop straight down} [gesturing his pencil downwards], {\em and harder for them to...} [looks at the problem sheet], {\em is that right? } 

    [Pauses for 20 seconds while looking back and forth between the problem and the equation sheet.] 

    {\em  Since its angular momentum, they will have... }[draws and erases a diagram and reads the problem statement again. Pauses for 17 seconds] 
\end{quote}

\subsubsection{Matthew's construction of a qualitative mental model}

The transcript till this point reflects Matthew's construction of a qualitative mental model. Matthew makes an incomplete reference to the rider's free fall (``[...]{\em harder for them to...''}) by abruptly interrupting his statement with the question ``{\em is that right?}" After a pause of 20 seconds, he refers to the rider's rotational motion through the statement ``{\em Since its angular momentum...}''. This shift in arguments, from the rider's linear to their rotational motion, coupled with the incomplete and uncertain nature of the associated arguments suggests Matthew's construction of a mental model~\cite{norman2014some}. Additionally, the downward-gesturing of his pencil during his reference to the rider's free fall reflects his mental imagery about the rider's spatial motion under the influence of gravity. Such physical gestures mimicking spatial motion of objects have also been associated with students' mental modeling~\cite{schwartz1996shuttling,goldin2005our}.

\subsubsection{Matthew's Assembling of a Knowledge Framework and Noticing of an Inconsistency}

On the sensemaking front, the same segment reflects the first two stages (Stage 0 and Stage 1) of the Sensemaking Epistemic Game. Matthew's initial moves -- highlighting the relevant information in the problem statement and emphasizing the rider's mass -- corresponds to his assembling of knowledge about the Gravitron task. Investigations on physics problem solving have noted the role of problem statements in activating students' prior knowledge~\cite{hammer2005resources,ACER,shar2020student}.

Matthew's argumentative shift in the remaining part of the transcript highlights his entry condition into the Sensemaking Epistemic Game by noticing a gap in his knowledge. The perceived gap is between his existing knowledge on linear motion (ease of free fall by a heavier person) and the required knowledge on rotational motion (angular momentum) to determine the rider's status inside the Gravitron. The noticing is evidenced by the  puzzling question ``{\em is that right?}"  sandwiched between the two arguments.  Odden and Russ~\cite{odden2019vexing} note that such questions reflecting the essence of `something not being right' mark students' entry into the Sensemaking Epistemic Game.

Summarizing the episode, Matthew's priming on the task-related information and his argumentative shift from linear to rotational motion reflect his construction of a mental model in addition to his navigation through the first two stages of the Sensemaking Epistemic Game.

\subsection{Episode II}

Matthew then argues that for a rider to be suspended inside the Gravitron, the normal force should exceed the downward pull of gravity.

\begin{quote}
 {\em So you need... since the rider is gonna be thrown into the wall, you need a normal force that exceeds the downward force of gravity.} [Pauses for 38 seconds.]    
\end{quote}

The prolonged pause makes the interviewer intervene with the question, ``{\em What are you thinking?}" to which he replies as: 
\begin{quote}
{\em I am just trying to remember how to solve the problem.}
\end{quote}

Constructing a force diagram, Matthew represents the forces of friction and gravity (Figure~\ref{fig:Matthew-attempt}), and calculates the magnitude of the rider's centripetal acceleration by assuming that the Gravitron's walls are perpendicular to the ground.

\begin{quote}
    {\em  I know there is... in the free body diagram, there is the downward force due to gravity} [simultaneously draws the force diagram and indicates gravity as $F_g$], {\em and then the normal force from being thrown into the wall} [draws the normal force and indicates frictional force as $F_f$], {\em and then that creates the force of friction. So, you need to solve for centripetal acceleration.} 
    
    [By looking at the equation sheet] {\em We are given omega, so we can solve for, solve for centripetal acceleration. And then we can do F equals ma. Assuming that the wall is at ninety degree to the ground. Then... this should be correct. The centripetal acceleration is r times omega squared. So $4.6$ times... yeah that's in radians per second. That's proper? And that's [types in calculator] $13.8$.}
  \end{quote}

Reiterating the same assumption, he goes on to equate the denoted forces of friction and gravity. With the help of the provided equation sheet, the formulated equation is then simplified into an algebraic inequality expressed in terms of the centripetal acceleration (13.8), the coefficient of static friction (0.2), and the acceleration due to gravity (9.8) as shown in Figure~\ref{fig:Matthew-attempt}. 

\begin{quote}
[Scribbling on the solution sheet] {\em Assuming that it's ninety degree, all of their... all of that will be converted into frictional force }[looking at the equation sheet]. {\em Yeah... that's the normal force.. so F equals ma. Yeah, F equals } [Pauses for 15 seconds, looks at the equation sheet]. {\em Is there a friction section here? Yeah } [finds the required section]. {\em So its the normal force into the wall, the force of friction  has to equal the force of gravity, and both have mass, and this mass is negligible, so you have just... a equals... and then this a has to be bigger than or equal to} [erases].  

{\em So, the force of friction is normal force times mu equals force of gravity. Normal force in this equation is the angular acceleration times mass. For F equals ma times mu, equals mass times gravitational acceleration, and then both have mass} [cancels `m' on both sides of the equation]. {\em So yeah $13.8$ times the $0.2$, coefficient of static friction, and that has to be greater than or equal to 9.8 for the gravitron to be able to hold people up.} [Uses calculator]
\end{quote}

We are cautious to interpret most of the first thirteen lines (the first segment) of the above excerpt (``{\em Assuming that it's ninety degree [...] bigger than or equal to.}'') since Matthew utters these arguments while scribbling on the solution sheet which he later erased (as noted in the transcript). Because of the lack of evidence in corroborating the student's verbal arguments with his written solution, we cannot ascertain the intended meaning of these these lines.

\begin{table*}
\begin{center}
\caption{Summary of the features of the epistemic games in Matthew's and Ken's approaches at the Gravitron task. Both the approaches share the same target epistemic form of making a claim about the rider's status inside the Gravitron. The target epistemic form is further constrained by only two possibilities: either or not the riders would fall of the Gravitron's walls.\label{tab:epistemic-game-features}}

\begin{ruledtabular}
\begin{tabular}{p{2.5cm} p{7cm} p{7cm}}

Features  & Matthew's approach  & Ken's approach \\
\hline

Entry condition & Noticing of a gap between his existing knowledge on linear motion and expected knowledge on rotational motion. & Noticing of a gap in his understanding on the interplay of the forces in holding up the Gravitron's rider. \\

Moves & Highlights the information from the problem statement.   & Identifies the information from the problem statement. \\

        & Argues about linear and rotational motion of the Gravitron's rider.                &  Constructs force diagram.      \\
        
        & Constructs a force diagram and `mathematizes' the diagram-based arguments.                    & Reasons about the association between the indicated forces in the constructed diagram.           \\
        
        & Interprets the physical implication of the formulated mathematical inequality to make a claim about the Gravitron's rider. & Quits the problem-solving exercise \\
        
Exit condition & Concludes that the riders would slide off the Gravitron's walls under the given conditions.   & Quits the problem without making any claim.  \\

\hline
\end{tabular}
\end{ruledtabular}
\end{center}
\end{table*}
 
\subsubsection{Matthew's engagement with the Denotative Function and the Demonstration components of modeling}
\label{subsec:matthew-episode2}

The current episode's transcript reflects the modeling's Denotative Function and the Demonstration components in his approach. The student focuses on the rider-Gravitron system (target system), and represents it through a force diagram highlighting the forces of friction ($F_f$) and gravity ($F_g$). Matthew's moves to represent the target system through a force diagram marks the modeling's Denotative Function in his approach. Furthermore, Matthew's moves to equate the denoted forces (i.e., mathematically relating the denoted elements of the representation), and simplifying the formulated equation into the algebraic inequality reflect the modeling's Demonstration component in his reasoning.  

\subsubsection{Matthew's Generation of an Explanation}

Through the lens of sensemaking, the same segment marks Matthew's  response to the perceived need to account for the rider's rotational motion as observed in the previous episode (Section~\ref{subsec:matthew-episode1}). Matthew generates an explanation by constructing the force diagram, calculating the centripetal acceleration, and determining the inequality (in terms of the centripetal acceleration). These moves represent Matthew's navigation through the explanation generation stage (Stage 2) of the Sensemaking Epistemic Game. 

In summary, Matthew's moves in this episode correspond to his engagement with the Denotative Function and the Demonstration components of modeling. The same moves also reflect his navigation through the explanation generation phase of the sensemaking process. 

\subsection{Episode III}
\label{subsec:matthew-episode-III}

In the end, Matthew interprets the result of the obtained algebraic inequality and concludes that riders would fall off the Gravitron's wall under the given conditions. He further goes on to claim that the Gravitron needs to be spinning at a rate of 49 radians per second for people to hold up.      

\begin{quote}
{\em  Yeah, what I found is everyone is gonna slide because it's not spinning fast enough. And it need to be spinning}  [uses calculator] {\em 49~radians~per~second for people to hold up on the wall.}    
\end{quote}

As a clarifying note, we do not have evidence (either in Matthew's verbal arguments or in his written solution) reflecting the reasoning which guides the calculation of the required angular velocity of the Gravitron (49 radians per second). Consequently, the following analysis will not entail any interpretations about the same, except for acknowledging the statement. 

\subsubsection{Matthew's engagement with the Inferential Function component of modeling}

Matthew's final move of interpreting the physical implication of algebraic inequality through the lens of the rider-Gravitron system highlights the Inferential Function feature of modeling in his approach. The student transitions from the `model-world' back into the `physical world' by mapping the theoretical result (inequality) obtained in the Demonstration stage onto the target system. This mapping assists Matthew in making the required prediction thereby concluding the modeling process.

\subsubsection{Matthew's Resolution of the perceived gap}

Matthew's concluding statement also marks his exit from the sensemaking process and thus the Resolution stage of the Sensemaking Epistemic Game. He generates a coherent (mathematical) explanation by resolving his perceived need to account for the rider's rotational motion in making the required claim. He concludes clearly that riders would slide off the Gravitron's wall under the specified conditions, and further goes on to state that riders would need to be spinning at 49 radians per second to stay put. This decisive claim marks Matthew's exit condition from the Sensemaking Epistemic Game. Table~\ref{tab:epistemic-game-features} ('Matthew's approach' column) further summarizes features of epistemic games in Matthew's approach at the Gravitron task.

In summary, the concluding phase of Matthew's attempt at the Gravitron task simultaneously reflects the Inferential Function component of modeling and the Resolution stage of the Sensemaking Epistemic Game. Figure~\ref{fig:participants-summary} (a) summarizes the relative overlap between the elements of modeling and sensemaking processes in Matthew's solution approach. 

In the following section, we present another attempt at the same Gravitron task from a student named Ken (pseudonym). Unlike Matthew, Ken quits the problem-solving exercise upon making several unsuccessful bids. Similar to the analysis demonstrated in this section, we simultaneously examine Ken's problem-solving moves through the lens of modeling and sensemaking in addition to noting their relative overlap. 

\begin{figure*}
    \centering
    \includegraphics[scale=0.8]{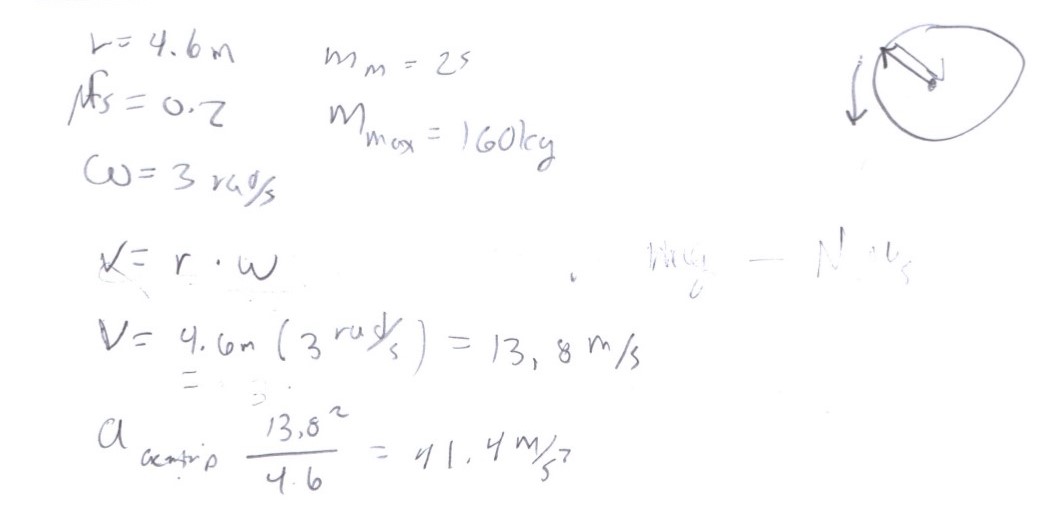}
    \caption{Ken's attempt to the Gravitron Problem}
    \label{fig:Ken attempt}
\end{figure*}

 \section{Ken's attempt at the Gravitron task:}
\label{sec:ken-attempt}

Upon pooling the contextual information, Ken represents the rider-Gravitron system (as shown in Figure~\ref{fig:Ken attempt}). While reasoning about the constructed representation, Ken suggests to compare the involved forces in order to make the required claim. Unable to relate the forces, Ken discontinues his attempt, and proceeds to solve other problems from the set. Revisiting the Gravitron task, he calculates the riders' centripetal acceleration and finally abandons his pursuit. In the rest of this section, we detail Ken's attempt along with analyzing his problem-solving moves through the lens of modeling and sensemaking. Figure~\ref{fig:participants-summary} (b) summarizes Ken's moves categorized into the components of modeling and stages of the Sensemaking Epistemic Game. 

\subsection{Episode I}
\label{subsec:ken-episode-I}

Ken begins by noting down the `given' quantities from the problem statement and refers to the riders' free fall through a hand gesture.

\begin{quote}
    {\em So, just like off the side, its kinda like write down the stuff it gives for later use}. 
    [Reads the problem statement for 12 seconds, and writes down the magnitudes of radius, coefficient of static friction, minimum angular velocity, minimum and maximum mass]. 
    
    {\em Okay, so ah} [looks at the equation sheet and reads the problem statement]. {\em I guess like in order to understand for them not to go down} [gestures his index finger downwards], {\em ...umm..., the force... uh.} 
\end{quote}

\subsubsection{Ken's construction of a qualitative mental model}

The above transcript reflects Ken's construction of a qualitative mental model. Ken tries to articulate a condition through a hand gesture that prevents the Gravitron's riders from slipping down. He however does not complete the argument which is evidenced by the phrases ``{\em umm... the force... uh}''. The incompleteness of Ken's arguments along with the representational gesturing of the riders' spatial motion reflects his engagement with a qualitative mental model~\cite{schwartz1996shuttling,goldin2005our}.

\subsubsection{Ken's Assembling of Knowledge Framework}

From the sensemaking perspective, the same segment marks Ken's assembling of his knowledge framework (Stage 0 of the Sensemaking Epistemic Game) evidenced by pooling of the relevant contextual information.

\subsection{Episode II}
\label{subsec:ken-episode-II}

Ken proceeds with construction of a representation as shown in Figure~\ref{fig:Ken attempt}. He then suggests to compare the involved forces after identifying one of the indicated arrows as the force preventing the riders from slipping down. He then refers to angular acceleration amid intermittent pauses.

\begin{quote}
{\em Okay so, I will draw the picture here} [draws a circle, two oppositely directed arrows and a vertically downward arrow. Reads the problem statement for 8 seconds] {\em And then this is...} [pointing at the outward arrow] {\em Uh... so this force has to keep them from falling down and then so, we need to find ah... compare, like the forces?} [Points at the diagram in the problem statement, and refers to the equation sheet for 17 seconds.]

{\em Okay, so, my equation sheet is uhh, that alpha is delta omega by delta t and then umm...} [Pauses for 12 seconds, and looks at the problem statement and the equation sheet].
\end{quote}

The recurring pauses makes the interviewer intervene with the question ``{\em What are you thinking?}" to which Ken replies that he is trying to relate the information from the equation sheet to the given context.

\begin{quote}
{\em Uhh, I am trying to see how I can relate uh, like what information I got to, just to see like what other information I think I kinda like to obtain to hopefully find like uh forces. I don't know if that makes any sense. Uh} [Looking towards equation sheet and pauses for 20~seconds]. 

{\em I think.. Yeah, what I am trying to think is how I am gonna relate this} [pointing at the Gravitron's image from the problem statement] {\em to forces} [pointing at the equation sheet]. [Refers the equation sheet for 15 seconds] {\em Okay, so I did not find anything there.}
\end{quote}


\subsubsection{Ken's engagement with the Denotative Function component of modeling}

This episode reflects the modeling's Denotative Function in Ken's reasoning. Ken constructs a representation by associating two radially opposite arrows and a downward arrow with forces experienced by the Gravitron's rider (``{\em So this force has to}[...]"). Through this representation, Ken explicitly highlights the rider-Gravitron system as the target system, and thus transitions from the physical-world to the model-world. In addition, his reply to the interviewer's question on trying to relate the information from the equation sheet to `{\em find forces}' suggests his attempt to associate the indicated arrows with the relevant forces. This feature of highlighting the target system through a representation, and associating features of the system (forces on the rider) through elements of the representation (arrows) marks the modeling's Denotative Function in Ken's approach.

\subsubsection{Ken's Noticing of an Inconsistency}

On the sensemaking front, the same segment reflects Ken's entry into the Sensemaking Epistemic Game through noticing of an inconsistency in his understanding of how the forces sustain the Gravitron's rider. The noticing is evidenced by (i) his suggestion to compare the forces, (ii) lack of reference to specific forces either in his verbal arguments or in his written solution, and (iii) an explicit mention of trying to relate the forces from the equation sheet (``{\em Yeah, what I am trying to think [...]}").

Summarizing the episode, Ken's construction of the representation, and his attempt to relate the forces from the equation sheet to the representation correspond to his engagement with the modeling's Denotative Function. The same moves also evidence his noticing of an inconsistency in his understanding on the interplay of the forces in holding up the Gravitron's rider.

\subsection{Episode III}
\label{subsec:ken-episode-III}

Struggling to locate clues from the equation sheet, Ken proceeds by reasoning about the forces acting on the rider. Noting static friction to oppose the rider's weight along the $y$-direction, he refers to the relationship between friction and the normal reaction. This relationship cues Ken on calculating centripetal acceleration. Unable to reason any further, he acknowledges of being `stuck' on the task, and mentions of revisiting it upon solving the remaining problems from the set.

\begin{quote}
 {\em Okay so, in the y direction, we have, will start with the max weight. So you have weight of person putting it down, and then static friction, would be opposing it slipping so, and static friction is normal force times $\mu_s$. And then normal force is equal to...} [looks at the equation sheet].  {\em Okay, so we need to find alpha} [pauses for 20~seconds while looking back and forth between equation sheet and problem statement]. {\em So....} [pauses for 10 seconds].

 {\em Yeah, I am pretty stuck on this right now. Just gonna, I will just come back to that. There are no points involved } [smiles and turns the page for the next problem].
\end{quote}

\subsubsection{Ken's engagement with the Demonstration component of modeling}

Ken's moves in this episode exemplify the modeling's Demonstration component. Having denoted forces as arrows in the representation, Ken identifies weight, friction, and the normal force as the forces experienced by the Gravitron's rider. He notes the force of friction to oppose the rider's weight, thereby tying the two forces together. He then connects the friction and the normal force by citing the mathematical relation between them. Consequently, Ken's attempt to relate the identified forces reflects the Demonstration component of modeling in his reasoning. 

\subsubsection{Ken's Generation of an Explanation}

On the sensemaking front, Ken's reasoning about the forces marks his navigation through the penultimate stage of the Sensemaking Epistemic Game. In response to the noticed inconsistency on the interplay of the forces on the Gravitron's rider, Ken generates an explanation by identifying the relevant forces at play and describing their interrelationships. He intuitively argues that the rider's downward motion would be opposed by the force of friction. He then invokes his formal knowledge by expressing friction as the product of normal force and the coefficient of static friction. Thus, Ken generates an explanation by blending his intuitive and formal knowledge to relate the identified forces.

Upon attempting the remaining problems, Ken returns to the Gravitron task and calculates the magnitude of centripetal acceleration. Noting the centripetal force to keep the `{\em kid on the ride}', he tries identifying the counter-force that works in opposition to the centripetal force.

\begin{quote}
{\em I will try this one real quick. Okay.} [Looks at the equation sheet and erases what was earlier written]. {\em So, $r$ times $w$. Radius is $4.6$ times $3$} [Uses calculator]. {\em And then} [looks at the equation sheet]. {\em Yeah okay. So, you can find centripetal acceleration by squaring the... and dividing it by $r$} [uses calculator]. {\em So its $44.1$~meters~per~second-squared. And this acceleration has to do like with the force that keeps the kid on the ride [while pointing at the radially inward arrow in the diagram]. And then okay.} [Pauses for 42~seconds] {\em How do I...?} [Points at the radially outward arrow in the diagram and pauses for 10~seconds]. {\em So, there is a force pushing back on them}. [Pauses for 22~seconds].
\end{quote}

Ken continues to relate the forces, this time by focusing on the radial arrows indicated in his representation. He notes the radially inward arrow as the centripetal force that holds the rider inside the Gravitron (``{\em [...] the force that keeps the kid on the ride.}''). He then tries to reason about the oppositely directed arrow acting in opposition to the centripetal force. Ken's attempt to contrast the indicated arrows in his representation highlights his continued engagement with the Demonstration component of modeling.    

Ken's determination of the centripetal acceleration and reasoning about the two forces also reflect his sustained efforts to generate explanations on the sensemaking front.

In addition, this segment also corresponds to Ken noticing an additional inconsistency on fleshing out the `pushing back' force acting on the rider. This is evident from the question ``{\em How do I...?}" as such vexing questions have been noted to reflect students noticing additional inconsistencies during sensemaking~\cite{odden2019vexing}. Ken's noticing of additional discrepancies while generating explanations is also consistent with the trajectory of the sensemaking process as highlighted by the Sensemaking Epistemic Game framework~\cite{sensemakinggame}.

Unable to reason conceptually on the `push back' force, Ken resorts to his memory on trying to recall a solution approach. Unable to recall a prior solution, he finally abandons his pursuit on the Gravitron task.  

\begin{quote}
{\em Umm... I am trying to like, umm, I am trying to recall, like the similar problem I did. There is a force...} [pauses for 22 seconds] {\em Well, I think that's the best I could do so far.}
\end{quote}

This last segment of the transcript reflects Ken's unsuccessful attempt at the Demonstration component of modeling. In pursuit of determining the push-back force, Ken seeks to recall a strategy from his memory. Unable to do so, he quits the problem solving exercise.  

From the sensemaking perspective, this segment marks Ken's abrupt exit from the epistemic game. Without a satisfactory explanation on the interplay between the forces in holding up Gravitron's rider, Ken gives up on his attempt after not being able to recall a solution approach from his memory. Table~\ref{tab:epistemic-game-features} (column 3) presents the features of epistemic games in Ken's attempt at the Gravitron task. 

Summarizing the episode, Ken identifies weight, frictional and normal forces and relates them in the Gravitron context through his intuitive, and curricular knowledge. Struggling to make inroads, he temporarily abandons the task and revisits it. He then seeks to determine the `push-back' force which acts in opposition to the centripetal force. Unable to make a breakthrough, the student abandons his solution pursuit. From the modeling perspective, Ken's moves in this episode reflect his struggle to relate the denoted forces in his representation, i.e., the struggle to engage in the modeling's Demonstration component. On the sensemaking front, the same moves reflect his struggle to generate a coherent explanation for his perceived gap about the forces acting on the Gravitron's rider.

\section{Discussion}
\label{sec:Discussion}

In order to probe how modeling and sensemaking intertwine, we analyzed two case studies of physics problem-solving through the lens of mental modeling, the DFDIF components, and the stages of the Sensemaking Epistemic Game (Figure~\ref{fig:participants-summary}). We observe that modeling -- construction of mental models and engagement with DFDIF components --  entails navigation through the stages of the Sensemaking Epistemic Game. Even though the components of modeling overlap with the epistemic game's stages, we do not find a one-to-one relation between the elements of the two processes. We also observe that barriers experienced in engaging with the modeling's DFDIF components can inhibit navigation through the stages of the Sensemaking Epistemic Game. These arguments are detailed below.    

\subsection{Modeling involves navigation through the stages of the Sensemaking Epistemic Game}

Figure~\ref{fig:participants-summary}, particularly (a), indicates that the construction of a mental model and engagement with the DFDIF components involve navigation through the stages of the Sensemaking Epistemic Game. Additionally, the figure also highlights the co-occurrence of the assembling of knowledge framework stage of the epistemic game (Stage 0) with the construction of a mental model; the explanation-generation stage (Stage 2) with the Demonstration; and the Resolution stage (Stage 3) with the modeling's Inferential Function.   

In the initial phases of problem-solving, both participants primed on the task-related information as part of constructing their mental models. While Matthew annotated the provided information, Ken wrote out the same explicitly. The assembling of the task-related information in both cases contributed to the participants' interaction with their mental models. Our observation on the overlap between mental modeling  and assembling of contextual information is consistent with the current literature, which has noted mental modeling as a key feature of making sense of the task~\cite{sands2021modeling} using the essential pieces of information from the problem statement~\cite{cortes2021makes,roberts2000strategies}. 

We also note the alignment of the  explanation generation during sensemaking (Stage 2) with the modeling's Demonstration feature. Both Matthew and Ken generated explanations in response to their perceived gap by mathematically relating the denoted forces in their respective representations. The alignment of the explanation generation stage with the Demonstration component comes as no surprise as numerous studies have noted sensemaking to involve coordination between multiple representations~\cite{emighstudent,maloney2015teaching}. {\em Demonstration}, particularly while making sense of physics problems, can involve making effective transitions between various forms of representations such as Free Body Diagrams, graphs, equations, etc.\ ~\cite{gire2015structural,susac2017graphical}. 

Lastly, in case of Matthew, the Resolution stage of the epistemic game (since Ken did not reach this stage) aligned with the modeling's Inferential Function. Matthew concluded his sensemaking by interpreting the result of the mathematical inequality, i.e., by transitioning from the model-world to the physical-world. Our observation on the overlap between the results' interpretation (Inferential Function) as the concluding part of the problem solving process (Resolution) is also congruent with the existing findings on students' use of mathematics in physics~\cite{caballero2015unpacking,ACER,sands2021modeling}.  

Even though the components of modeling co-occur with stages of the Sensemaking Epistemic Game, we do not observe a one-to-one correspondence between all the elements of the two processes. This primarily stems from the dynamic nature of noticing inconsistency during sensemaking. While Matthew noticed a discrepancy in his knowledge during mental modeling (Section~\ref{subsec:matthew-episode1} and Figure~\ref{fig:participants-summary} (a)), for Ken it was associated with the Denotative Function and while {\em Demonstrating} the internal consistency in his representation (Section~\ref{subsec:ken-episode-II}~\&~\ref{subsec:ken-episode-III} and Figure~\ref{fig:participants-summary}(b)). 

We argue that the association of modeling as a process of making sense of a phenomenon~\cite{chen2022epistemic,russ2017intertwining,schwarz2009developing,schwarz2017helping,k-12framework,passmore2014models,sands2021modeling} can be restated more precisely as: \textit{modeling as a process of navigating through the stages of the Sensemaking Epistemic Game when reasoning about a phenomenon.} The contemporary literature has also noted several sensemaking features, such as priming of prior knowledge, and generating explanations, as part of the modeling process~\cite{chen2022epistemic}. Our findings complement this observation by noting that these features co-occur vis-\`a-vis engagement with mental models and with the Demonstration feature of modeling. 

\subsection{Struggling to engage with one or more components of modeling can inhibit sensemaking}

The analysis of Ken's case further reveals that barriers experienced in modeling can impede sensemaking of the given situation. From Figure~\ref{fig:participants-summary} (b), we observe that the remaining facets of sensemaking (the Resolution stage) and the modeling processes (the Inferential Function component) are not evidenced as Ken chooses to abandon his attempt at the task. Despite representing the rider-Gravitron system, and reasoning about the forces acting on the Gravitron's rider, Ken’s choice to give up on his approach stemmed primarily from his repeated unsuccessful attempts to relate the two oppositely directed arrows in his representation. Upon noting the centripetal force (the radially inward arrow in his representation), Ken struggled to reason about the `push-back' force (the radially outward arrow) which acted in opposition to the identified centripetal force. From the modeling perspective, Ken's struggle tends to be associated with relating the elements (forces) of the constructed representation, i.e., the Demonstration component of modeling. The hindrance in the modeling's Demonstration component resulted in the lack of a satisfactory explanation to his perceived gap, which in turn nudged him to quit the sensemaking process.

We emphasize that Ken's decision to give up on his attempt may not have been caused {\em solely} by the modeling barriers. The lack of conceptual coherence in his arguments, or his `working-model' of the Gravitron may also have played a crucial role in nudging him to quit sensemaking. However, we prioritize modeling barriers over other factors since modeling or sensemaking can occur independently of conceptual coherence. Other participants in our data made the requisite claim upon making sense of or modeling the Gravitron scenario, albeit their conceptual correctness can be contested. Furthermore, our observation on modeling barriers impeding sensemaking addresses the contemporary literature's call to investigate the contextual factors which encourage students to give up on sensemaking~\cite{sensemakinggame}.

\section{Conclusion and Implications}
\label{sec:conclusion}

We analyzed two case studies of physics problem-solving in which two introductory students, Matthew and Ken, individually made sense of the Gravitron task (Figure~\ref{fig:Gravitron}) by modeling its context. The current work explicitly demonstrates the intertwining between sensemaking and modeling by employing the Sensemaking Epistemic Game~\cite{sensemakinggame}, mental modeling and the DFDIF account of modeling~\cite{dfdif}. Qualitative analysis of the participants' problem-solving reveals that modeling  -- constructing a mental model and engaging with the DFDIF components -- entails navigation through the four stages of the Sensemaking Epistemic Game. In addition, the analysis of Ken's approach reveals that barriers experienced in modeling can influence students to prematurely quit sensemaking.

Observations made in the current study can provide insights for instructors for making potential interventions when students tend to give up on sensemaking. For instance, a useful intervention in the case of Ken, who abandoned sensemaking upon struggling with modeling's Demonstration component, might have been to prompt him to reflect on how his identified forces -- gravity, friction and normal force -- connect to the denoted centripetal force. We suspect students can also experience barriers with other modeling features such as construction of mental models, and engagement with the Denotative Function, and the Inferential Function components (though not observed in our case studies). Students' initial mental models particularly while reasoning about a real-world context like the Gravitron, are associated with ideas about the functioning of their surrounding world. Potential interventions on mental modeling might include prompting students to explicitly articulate their intuitive ideas and discussing about how those ideas manifest as scientific principles in the given context. As for modeling's Denotative Function, the interventions can include explicitly making the ‘epistemic status’ of denoting physical quantities through algebraic symbols as an arbitrary choice often devoid of physical or mathematical implications (e.g., denoting Friction as `$F_f$' or `$f$' has no bearing on physics or mathematical arguments being made in a given context)~\cite{sirnoorkar2020towards,sirnoorkar2016students}. Concerning the Inferential Function, explicit emphasis about reflecting on the `physical meaning' of the obtained mathematical results, or modifying known equations in light of given physical conditions can assist students in sensemaking during problem solving. 


For researchers, our contextual operationalization of the Sensemaking Epistemic Game contributes a new theoretical perspective to analyze students' sensemaking on physics problems. Even though researchers have probed students' reasoning using the epistemic games construct, the focus has primarily been on the use of mathematics in physics~\cite{tuminaro2007elements} or `answermaking' during problem solving~\cite{chen2013epistemic}. Students' sensemaking using mathematical formalisms or `mathematical sensemaking' has been investigated through various constructs such as mediated cognition~\cite{gifford2020categorical}, blended processing~\cite{kuo2013students}, symbolic forms~\cite{sherin2001students}, etc.\ The current work adds the Sensemaking Epistemic Game to this list of frameworks employed to probe students' sensemaking during problem solving. And lastly, consistent with the current arguments in the modeling literature~\cite{gouvea2017models,passmore2014models}, this work takes the agent-based perspective to analyze students' modeling. That is, we investigate modeling through the cognitive interests and aims of the participant (models-for) rather than considering their mere representation of the concerned context (models-of)~\cite{gouvea2017models}.  

While our work affords instructors and researchers with insights, there are limitations to this study. The DFDIF account, which has been used as our modeling framework,  captures the `big-picture' of the modeling process by categorizing it into three broad activities: the Denotative Function, Demonstration, and Inferential Function. We note that modeling is a complex iterative process entailing activities such as making assumptions, approximations, etc.\ and our adopted framework does not capture these aspects. The arguments made in this study are based on the analysis of responses of two white, male students on a single problem. We acknowledge that a person's view of the outside world is informed by their positionality, and the claims made in this paper do not take this factor into account. Analysis of demographically diverse students' approaches to a wide range of problems would undoubtedly enrich the claims made in this paper. And lastly, while our data includes participants thinking aloud, there are moments especially during prolonged pauses which involve interventions from the interviewer asking participants to articulate their thoughts. Though an inherent limitation of the think-aloud protocol, the interviewer's interjection does have an impact on the participants' thought process.

As part of future work, we seek to extend our observations to identify the modeling-based task features which can promote sensemaking. Our observations on the association between the features of modeling and the stages of the Sensemaking Epistemic Game opens up avenues to investigate on the nature of target systems or contexts, modeling which can increase the likelihood of students' sensemaking. The findings also encourage us to look into the features of the {\em `Demonstration'}, especially in physics contexts, which can nudge students in connecting their everyday and curricular ideas. Lastly, the importance of the constructed inequality in Matthew's solution to both his modeling and sensemaking suggests that investigations exploring how the target epistemic form -- the end goal of the task -- may cue students to blend conceptual arguments with mathematical formalisms as another fruitful avenue of future research. Additionally, we also seek to investigate whether the converse of our second claim holds true, i.e., whether barriers experienced in sensemaking can inhibit modeling of the given context. Such explorations can further the pedagogical efforts in supporting students' modeling and sensemaking thereby making  classroom experiences more exploratory.

\section{Acknowledgements}
We would like to thank Katherine C. Ventura for creating the tests and conducting the interviews. Also thanks to Tor O.B. Odden, Rosemary Russ, Matthew Mikota, Brandi Lohman, and Hien Khong for their valuable insights. This work supported by the National Science Foundation under Grant Nos.\ 1726360 and 1725520.

\bibliography{ref}
\end{document}